# Engineering excitonic metal-insulator transitions in ultra-thin doped copper sulfides

Haiyang Chen[‡†], Yufeng Liu[‡†], Yashi Jiang[‡], Changcang Qiao[‡], Tao Zhang[‡], Jianyang Ding[⊥], Shuaishuai Yin[⊥], Zhengtai Liu[⊥], Zhenhua Chen[⊥], Yaobo Huang[⊥], Xiaolong Li[⊥], Hengxin Tan[‡], Jinfeng Jia[‡, §], Shiyong Wang[‡, §] *, Peng Chen[‡]*

[‡]Key Laboratory of Artificial Structures and Quantum Control (Ministry of Education), Tsung-Dao Lee Institute, Shanghai Center for Complex Physics, School of Physics and Astronomy, Shanghai Jiao Tong University, Shanghai 200240, China.

[⊥]Shanghai Synchrotron Radiation Facility, Shanghai Advanced Research Institute, Chinese Academy of Sciences, 201204 Shanghai, China

[§]Hefei National Laboratory, Hefei 230088, China

†These authors contributed equally to this work.

## ABSTRACT

Delicate engineering of the bands remains challenging due to complex electronic, structural, and compositional interplay. We demonstrate the formation of distinct metallic and insulating ground states in ultra-thin copper sulfide films by effectively tuning the band structure via changing the composition of Cu and S in the system. Using angle-resolved photoemission spectroscopy (ARPES), we observed a continuous band renormalization and opening of a full gap at low temperatures over a wide range of doping. The electronic origin of this metal-insulator transition is supported by scanning tunneling microscopy (STM) and low energy electron diffraction (LEED) measurements, which show no indication of superlattice modulation and lattice symmetry breaking.

The evidence of excitonic insulating phase is further provided by carrier density dependent transitions, a combined effect of electron screening and Coulomb interaction strength. Our findings demonstrate tunability of the band structure of copper sulfides, allowing for new opportunities to study exotic quantum phases.

Many-body interactions hold significant importance in solids, allowing for the manifestation of collective excitations and macroscopic quantum phenomena.[1-3] One example is excitonic insulating phase originated from the coulomb interaction which gives rise to the spontaneous formation of bound electron-hole pairs.[4-8] Although success in realization of such a state in atomic double-layer system via electrically grating, conclusive evidence in a natural material remain challenging as considerable attention was paid to materials with interacting electron-hole pockets located at different momentums.[9-17] Condensation of excitons with finite-momentum will result in a charge density wave (CDW) which is difficult to distinguish from Peierls-type CDW induced by a lattice distortion. It is more promising to identify a condensed phase with zero-momentum excitons.[18,19] However, exciton density deeply relies on the relative energies of electron and hole pockets. Methods of delicate control of the band structures thus become necessary.

Copper sulfides are a family of chemical compounds and minerals with the formula $Cu_xS$ ($1 \leq x \leq 2$).[20-23] They are intensively studied for the potential applications in thin-film solar cells and optoelectronics.[24,25] Although the long-standing interest in $Cu_xS$, its band structure has not been well revealed by experiments. Because of the structural complexity, $Cu_xS$ has different phases. The stoichiometric compound CuS (covellite) is a *p*-type metal and $Cu_2S$ (chalcocite) is a *p*-type semiconductor, in which the bulk band gap determined from optical absorption measurements is ~1.1-1.4 eV depending on the variation from the stoichiometric composition.[22, 26–30] It is thus expected that the band gap of $Cu_xS$ can be tuned via changing the composition of Cu/S in the system. This can be realized by annealing CuS films at high temperatures (220-450 ºC) as CuS tends to decompose above 220 ºC.[31] The loss of sulfur atoms will lead to a compositional change and form a series of $Cu_xS$ with x in the range of [1, 2]. Furthermore, excitons have been shown around the onset of the band gap in $Cu_xS$,[30, 32] which can result in an excitonic instability in the low temperatures when the gap becomes small, making $Cu_xS$ an ideal platform for investigating the long-sought insulating state induced by spontaneous exciton condensation.

We start with growing ultrathin CuS films using molecular beam epitaxy (MBE) on a bilayer-graphene-terminated SiC substrate. ARPES spectra taken at 10 K show hole-like bands centered at the $\bar{\Gamma}$ point (Figure S1), in consistent with the first-principles calculated results.[22] These bands cross over the Fermi level, indicating the system is metallic. Annealing of CuS film at different temperatures results in a varied composition of Cu and S, as manifested in different band structures observed by ARPES (Figure S1). For example, annealing at 330 ºC gives rise to a less *p* type doped system. More occupied valence bands below the Fermi level emerges compared with the CuS, as indicated in the ARPES spectra in Figure 1f and Figure S1. Sharp reflection high-energy electron diffraction (RHEED) patterns (Figure 1a) reveal a high-quality and well-ordered film.

The scanning tunneling microscopy (STM) measurements reveal $Cu_xS$ islands with an uniform thickness distribution on top of graphene (Figure S2). A Fourier transform of the atomic topographic image (Figures 1c and 1d) shows sharp (1 × 1) hexagonal lattice peaks and the in-plane lattice constant is determined to be ~4.0 Å, a lattice constant that is close to that of $Cu_2S$ compared to the CuS compound (3.79 Å).[22, 25, 26] Importantly, there is no superlattice modulation or translation symmetry breaking observed in the STM measurements at 3.5 K (Figure S3). Multiple hole-like bands around the $\bar{\Gamma}$ points are clearly observed in the ARPES spectra (Figure 1f, Figure S4). The α band is more prominent in the *s*-polarized spectra because of the matrix element effects, indicating Cu $3d_{xy}$ orbitals dominate in this band (Figures S5 and S6).[13, 33] The γ band exhibits very weak intensity and merges with the β band at lower binding energies (Figure S7). The calculated band structure of thin $Cu_2S$ layers with high chalcocite structure results in the best agreement with the experiments (Figure 1e, Figure S8). The chemical nonstoichiometry leads to a slight shift of Fermi level and the system becomes metallic.[34]

To study the temperature dependent behavior of the $Cu_xS$, we take systematic scans of the bands along $\overline{\Gamma M}$ (Figure 2) as a function of temperature. Interestingly, a full gap opening around the $\bar{\Gamma}$ point is observed when the temperature is decreased to 10 K, demonstrating that the system becomes a semiconductor/insulator at low temperatures (Figure S9). The gap is homogeneous in the momentum space. This is distinct from materials like 1T-$TaS_2$, 2H-$NbSe_2$, or $VSe_2$, where the CDW gap opens predominantly at specific nesting vectors of the Fermi surface, leading to a strongly anisotropic gap distribution in momentum space.[35-37] Strong band renormalization is observed as the flat band top is developed around the α band and the spectral intensity of the β band is suppressed near the Fermi level (Figure S10 and Figure S11). Selected temperature-dependent *s*-polarized ARPES spectra for a $Cu_xS$ sample are shown in Figure 2a. The gap formation is illustrated by the symmetrized ARPES maps in Figure 2b. The energy distribution curves (EDCs) around the zone center shown in Figure 2c were symmetrized with respect to the Fermi level. By symmetrization, the effect of Fermi-Dirac distribution at high temperatures can be canceled out. A single peak at the Fermi level indicates the α band top crosses over the Fermi level at temperatures higher than 150 K. At lower temperatures, the observed two-peak structure indicates the opening of a gap around the Fermi level. The spectroscopic gap is extracted by fitting the symmetrized EDC with a phenomenological self-energy expression:[38]

$$A(\mathbf{k},\omega) = \frac{B(\mathbf{k})}{\pi}\frac{\mathrm{Im}\sum(\mathbf{k},\omega)}{[\omega-\epsilon(\mathbf{k})-\mathrm{Re}\sum(\mathbf{k},\omega)]^2+[\mathrm{Im}\sum(\mathbf{k},\omega)]^2} \quad (1)$$

$$\sum(\mathbf{k},\omega) = -i\Gamma_1 + \frac{\Delta^2}{[\omega+\epsilon(\mathbf{k})+i\Gamma_0]} \quad (2)$$

where $A(\mathbf{k},\omega)$ is the spectral function, $B(\mathbf{k})$ is the corresponding weight, $\sum(\mathbf{k},\omega)$ is the self-energy, and Δ represents the gap. Examples of the fit to the EDC are shown in Figure 2(d) and the gap is determined to be ~80 meV at 10 K. The square of the energy gap as a function of temperature follows a functional form described by a semi-phenomenological mean-field equation (red solid curve in Figure 2e). The fit yields a transition temperature of $T_C$ = 153 ± 3 K. A similar transition temperature is obtained using the gaps determined from leading-edge midpoint of EDCs (Figure S12). As ARPES probes the spatially averaged electronic structure of sample surfaces, the defects/disorder and thermal effects at higher temperatures smears out spectral features, leading to the residual intensity around the Fermi level. The constant-energy-contour ARPES maps in Figure S13 exhibit a clear hexagonal symmetry, indicating the azimuthal disorder has no discernible impact on the ARPES bands.

The above gap analysis is performed assuming the particle-hole symmetry for the systems that exhibit transitions at low temperature. We took STS scans to probe the density of states above the Fermi level. As shown in Figure 3, STS results at 3.5 K reveal a small band gap (2Δ = 87 meV) determined from the difference of the valence and conduction band energy for the interior of the $Cu_xS$ islands (annealed at 320 ºC), which is in excellent agreement with the ARPES result (Δ = 48 meV). The density of states of conduction and valence bands are symmetric around the Fermi level, indicating the determination of the spectroscopic gap is reasonable and suggesting the existence of electron-hole coherent interaction. A notable particle-hole asymmetry is presented in the samples without a phase transition, as shown in Figure S14, in which more STS data and corresponding ARPES spectra for $Cu_xS$ with different Cu/S ratio are included.

The *in-situ* LEED results on a $Cu_xS$ film annealed at 330 ºC ($T_C$ = 153 K) are shown in Figure 3b. Clear diffraction spots with a hexagonal-shape (marked with red circles) from $Cu_xS$ are observed at 10 K, which is in consistent with the structure determined by STM. The other set of weak diffraction spots marked with blue dashed circles are from the less preferred domains aligned with the underlying graphene. The rest spots are from the reconstructed $(6\sqrt{3} \times 6\sqrt{3})$ SiC surface and graphene.[39, 40] The detailed assignments of the diffraction spots are shown in Figure S15. Temperature dependent LEED data shows there are no significant changes associated with a structural transition, including emergence of new diffraction spots and notable shift of the spots across the transition. These measurements reveal the absence of any structural phase transition in the low temperatures. The observed band structure change by ARPES is purely electronic and in consistent with the excitonic insulator picture that the excitons condense at low temperatures, which minimizes the energy of $Cu_xS$ system by opening a gap.

Annealing the CuS films at a higher temperature or with a longer time results in a similar band structure but a larger gap in ARPES measurements. To quantify the evolution of band structure with annealing temperatures, we extract the carrier density from the Luttinger area of the Fermi surface at 200 K in the normal phase.[41] The carrier density of the sample annealed at 330 ºC is determined to be 1.4 x $10^{13}$ $cm^{-2}$ (hole density $p$) around the zone center and the $T_C$ is more than 3 times of the value from the films annealed at 320 ºC with a carrier density of 3.8 x $10^{13}$ $cm^{-2}$ ($T_C$ = 40 K). However, annealing at higher temperatures causes a reduction of the $T_C$ and further annealing leads to a semiconductor/insulator in the normal phase, which shows a different band structure as α, β, and γ bands become fully occupied and merge at the $\bar{\Gamma}$ point (Figure S1). There is no obvious band structure change with decreasing temperature to 10 K, indicating the system becomes a band insulator (Figure S16).

The observed carrier density dependent behavior is consistent with the scenario of the excitonic insulator, a result from weakly screened Coulomb interaction in a semimetal. The summarized results are shown in a phase diagram (Figure 4). The excitonic transitions emerge in the $Cu_xS$ within a specific compositional window, corresponding to x ~ 1.35–1.9. Annealing of thin CuS films has two effects on the band structure instead of a rigid shift of the chemical potential: less *p*-type carriers around the Fermi level caused by the loss of sulfur atoms and increase of the energy difference between valence and conduction bands. The former effect reduces the screening strength and benefits the formation of the excitons, as shown in the increased $T_C$ for the samples with a hole density between 1.4 and 4.6 x $10^{13}$ $cm^{-2}$. However, the latter effect gives rise to a weakened Coulomb interaction, which dominates for samples annealed at higher temperature than 330 ºC and results in a smaller exciton density and a lower $T_C$. Note that $T_C$ is below 10 K (measurement limit) with the carrier density around the 4.6 x $10^{13}$ $cm^{-2}$, but the transition will be eventually destroyed in the high carrier density regime with significantly strong screening strength. The excitonic insulating phase has been reported in several doped systems.[42]

In summary, copper sulfides over a wide range of doping provide a versatile platform to realize quantum states based on the interband electron-hole coherent interaction. Flexibility of the Cu/S composition and the corresponding tunability of the band structure is the key to make this system a unique case. The behavior of the metal-insulator transition with varied carrier densities is understood as a combined effect of electron screening and Coulomb interaction strength. Future study including terahertz spectroscopy with high vacuum is promising to unveil the low energy collective modes formed during the condensation for a complete understanding of the nature of the excitonic insulator.[43]

## METHODS

## Film growth

The synthesis of copper sulphide films on substrates of SiC was done in the integrated MBE/ARPES systems at the lab in Shanghai Jiao Tong University. SiC were flash-annealed for multiple cycles to form a well-ordered bilayer graphene on the surface. We grow CuS thin films epitaxially by co-evaporating high purity Cu and S from an electron-beam evaporator and a Knudsen effusion cell, respectively, while the substrate was maintained at room temperature. S flux was achieved by thermally decomposed pure $FeS_2$ compound at around 800 °C. The band structures and lattices are characterized by ARPES, STM, and Raman measurements. The growth process and thickness of the films was monitored by real-time RHEED, and the growth rate was set to 60 minutes per layer of CuS. Control of the composition of Cu and S was achieved by variation of the annealing temperature under the S flux and annealing time. In particular, when the annealing temperature reaches 450 °C, the S loss is maximum and the system is close to $Cu_2S$. We adopt a carrier-density scaling method to estimate composition x. We linearly scale x between the known endpoints ($x = 1$ for CuS, $x = 2$ for $Cu_2S$) based on the measured carrier density around the Fermi level. Using this approach, we estimate $x = 1.28$ for samples annealed at 320 ºC and $x = 1.79$ for those annealed at 350 ºC. ARPES spectra taken for a series of $Cu_xS$ annealed at different temperatures are shown in Figure S1. The band structure of the annealed $Cu_xS$ film can be restored to that of the CuS phase by re-annealing the sample at low temperatures (below 200 °C) under a sulfur-rich flux.

## ARPES measurements

After growth and each annealing, the films were transferred *in situ* to the ARPES system at the lab in Shanghai Jiao Tong University. For the synchrotron-based ARPES measurements, a series of $Cu_xS$ films with different compositions were prepared and transferred via a vacuum suitcase to beamline 03U and 09U at Shanghai Synchrotron Radiation Facility. ARPES measurements were performed at a base pressure of ~$5x10^{-11}$ mbar with in-laboratory He discharge lamp (He-I 21.2 eV) and 30-100 eV photons at synchrotron using Scienta DA30 analyzers. Energy resolution is 15-25 meV and angular resolution is around 0.2°. Each sample's crystallographic orientation was precisely determined from the symmetry of constant-energy-contour ARPES maps. The Fermi level is determined by fitting ARPES spectra from a polycrystalline gold sample.

## STM/STS measurements

$Cu_xS$ films with different compositions were transferred via a vacuum suitcase to the STM system. STS/STM were carried out using the CASAcme cryogen-free STM under low temperature (3.5 K) and ultrahigh vacuum conditions ($3 \times 10^{-10}$ mbar). The tungsten tips were calibrated against the surface state of a Cu (111) single crystal. A lock-in amplifier (521 Hz, 5 mV modulation) were used to acquire dI/dV spectra. The STM images were processed with WSxM software.

## Computational details from first principles

Due to the presence of significant Cu disorder in the $Cu_2S$ crystal structure, we considered several different Cu occupation configurations while maintaining the chemical formula $Cu_4S_2$ within a hexagonal unit cell. For each configuration, full structural relaxation was performed. By comparing the calculated electronic band structures with experimental ARPES data, we identified the structure that best matches the experimental observations. This optimized structure is characterized by lattice parameters a = 3.99 Å and c = 6.87 Å, and atomic Wyckoff positions Cu1 (0, 0, 0.26022), Cu2 (1/3, 2/3, 0.43313), and S (1/3, 2/3, 0.7707), corresponding to the space group P-3m1 (No. 164). Since ARPES experiments were performed on few-layer samples, we further extracted a single-unit-cell-thick bilayer from this bulk structure to calculate its band structure. This bilayer model exhibits even better agreement with the experimental data. Consequently, the bilayer results were adopted throughout the manuscript. The layers were treated as freestanding in the calculations, as the interfacial interaction is of the van der Waals type between films and the graphene substrate, resulting in a nearly decoupled $Cu_xS$ overlayer.

All first-principles calculations were carried out using the Vienna Ab initio Simulation Package (VASP),[44,45] employing the projector augmented-wave (PAW) method.[46] Exchange-correlation effects were treated within the generalized gradient approximation (GGA), using the Perdew–Burke–Ernzerhof (PBE) functional.[47] A plane-wave energy cutoff of 400 eV was used. For the bilayer $Cu_2S$ model, the Brillouin zone was sampled using an 8×8 Gamma-centered k-point mesh. Structural optimization was performed until the residual forces on all atoms were smaller than 1 meV/Å. The electronic band structures were computed without including spin–orbit coupling.

## Film structure characterization

STM and LEED measurements show that islands of $Cu_xS$ are aligned with the underlying bilayer graphene by 0 º and 30º. A line profile shows the height of an as-grown island is ~2 nm (Figure S2). As the in-plane lattice constant (~0.4 nm) is close to $Cu_2S$, we estimate the number of layers using the thickness of a $Cu_2S$ unit cell (~0.68 nm) and the value is obtained to be three layers.[26] The temperature dependent behavior of the band structure is similar in the films with thinner thickness, which can be obtained by reducing the Cu flux rate or the growth time. However, there will be smaller coverage and the ARPES spectra are blurry. Sharper and cleaner ARPES spectra are obtained in the thicker films, allowing us to extract more accurately the band positions and gaps.

## X-ray diffraction measurements

X-ray diffraction measurements were performed on the surface x-ray scattering station (BL02U2) at the Shanghai Synchrotron Radiation Facility. The energy of the incident x-ray radiation was chosen to be 9.8 keV. The scattered radiation was detected by an Eiger 500K area detector. The films were capped with amorphous Selenium to minimize the surface contamination. Selenium is selected over sulfur due to sulfur's higher vapor pressure at ambient temperature. The XRD results (Figure S17) confirm the lattice constants estimated from the STM characterization.

**ASSOCIATED CONTENT**

**Supporting Information**

The Supporting Information is available free of charge at Nano Letter online.

ARPES spectra for a series of $Cu_xS$ films; additional STM data; photon energy and polarization dependent ARPES; calculated band structure; constant-energy-contour maps; LEED result; X-ray diffraction data.

## AUTHOR INFORMATION

**Corresponding Authors**

*Email: shiyong.wang@sjtu.edu.cn;

pchen229@sjtu.edu.cn

**Notes**

The authors declare no competing financial interest.

## ACKNOWLEDGMENTS.

We thank Shengwei Jiang, Jing Wang, Xiaoyan Xu, Mingpu Qing, Yanghao Chan for helpful discussions. The work at Shanghai Jiao Tong University is supported by the Ministry of Science and Technology of China under Grant No. 2022YFA1402400, No. 2021YFE0194100, No. 2020YFA0309000, the Science and Technology Commission of Shanghai Municipality under Grant No. 21JC1403000 and 24PJA051, the National Natural Science Foundation of China (Grant No. 12374188, No. 12488101, No. 22325203, No. 92265105). P. C. and H. T. thanks the sponsorship from Yangyang Development Fund.

**Figure 1**. Film structure and electronic band structure of ultra-thin $Cu_xS$. (a) A RHEED pattern of thin $Cu_xS$ film taken at room temperature. (b) Core-level photoemission spectra taken with 110 eV photons. (c) An atomic topographic image of $Cu_xS$ film taken at 3.5 K. ($V_{Bias}$ = −1.5 V; $I_t$ = 100 pA). (d) A pattern derived from the Fourier transform of the image. (e) Calculated band structure of bilayer $Cu_2S$ with Cu vacancies ($Cu_{1.99}S$). The Cu-deficient composition was determined through integration of the density of states by aligning the calculated Fermi level with the experimental value. The calculated band structure of bilayer $Cu_2S$ serves as a reference for that of $Cu_{1.99}S$. (f) ARPES maps taken with *s* polarized light for $Cu_xS$ annealed at 330 ºC in the normal phase at 200 K. The band dispersions (red dots) extracted by fitting to the EDCs and MDCs are superimposed on the top of the spectrum.

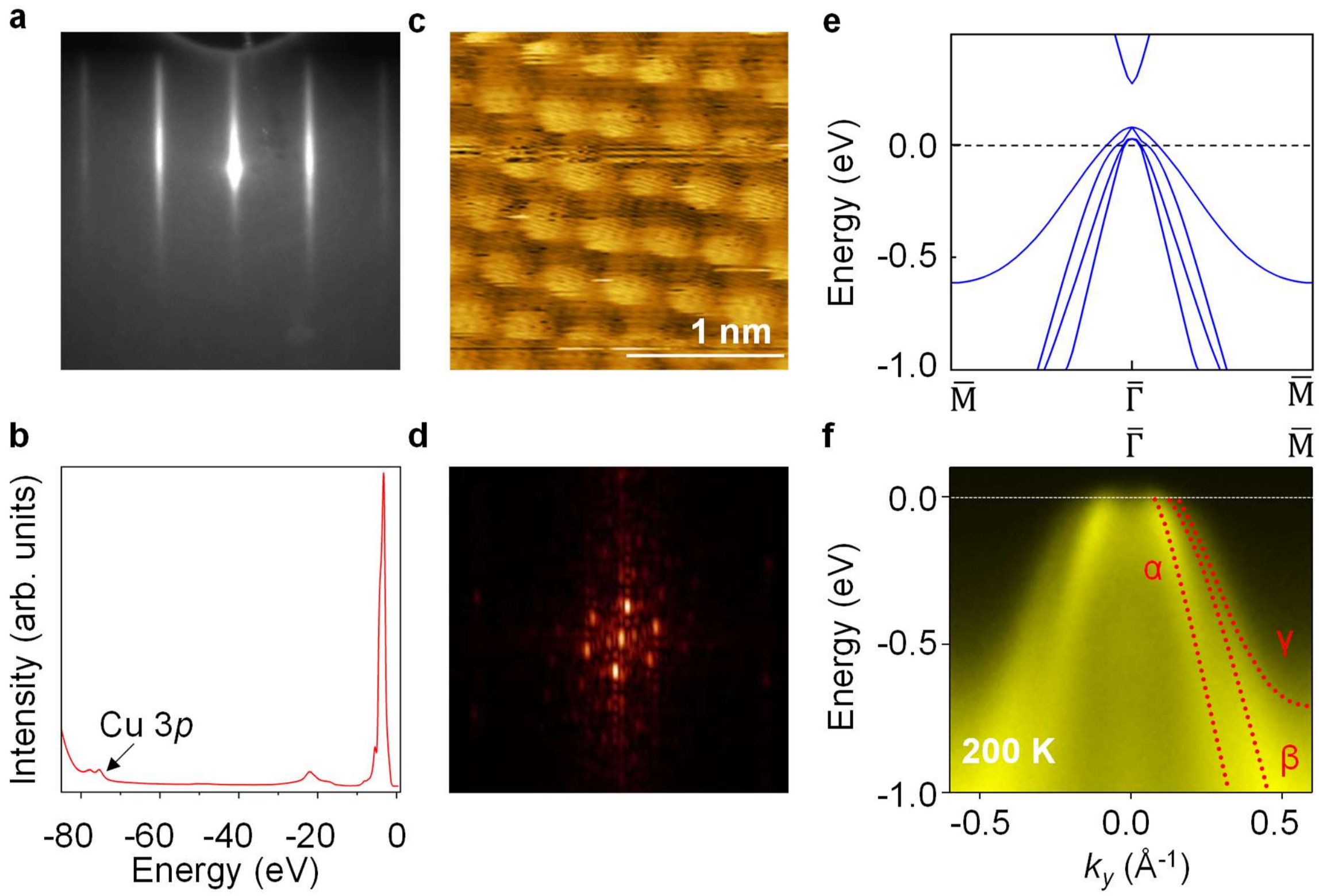

**Figure 2**. Temperature dependence of the band structure and the energy gap. (a) ARPES spectra reveal the development of the flat top of α band when the temperature is decreased from 200 to 10 K. The data were taken with 28 eV *s*-polarized photons. The red arrow indicates the momentum position that the α band crosses the Fermi level. (b) Corresponding symmetrized ARPES maps show the evolution of the gap around the zone center with temperature. (c) EDCs around the zone center indicated by the red arrow in (a) at selected temperatures, the leading-edge midpoints are demonstrated by a black arrow. (d) Symmetrized EDCs near the zone center at selected temperatures between 10 and 300 K. By symmetrization, the effect of Fermi-Dirac distribution at high temperatures can be canceled out. The solid lines are the fitting results. (e) The extracted temperature dependence of the square of the energy gap. The red curve is a fit using a BCS-type mean-field equation. Transition temperature $T_C$ is labeled.

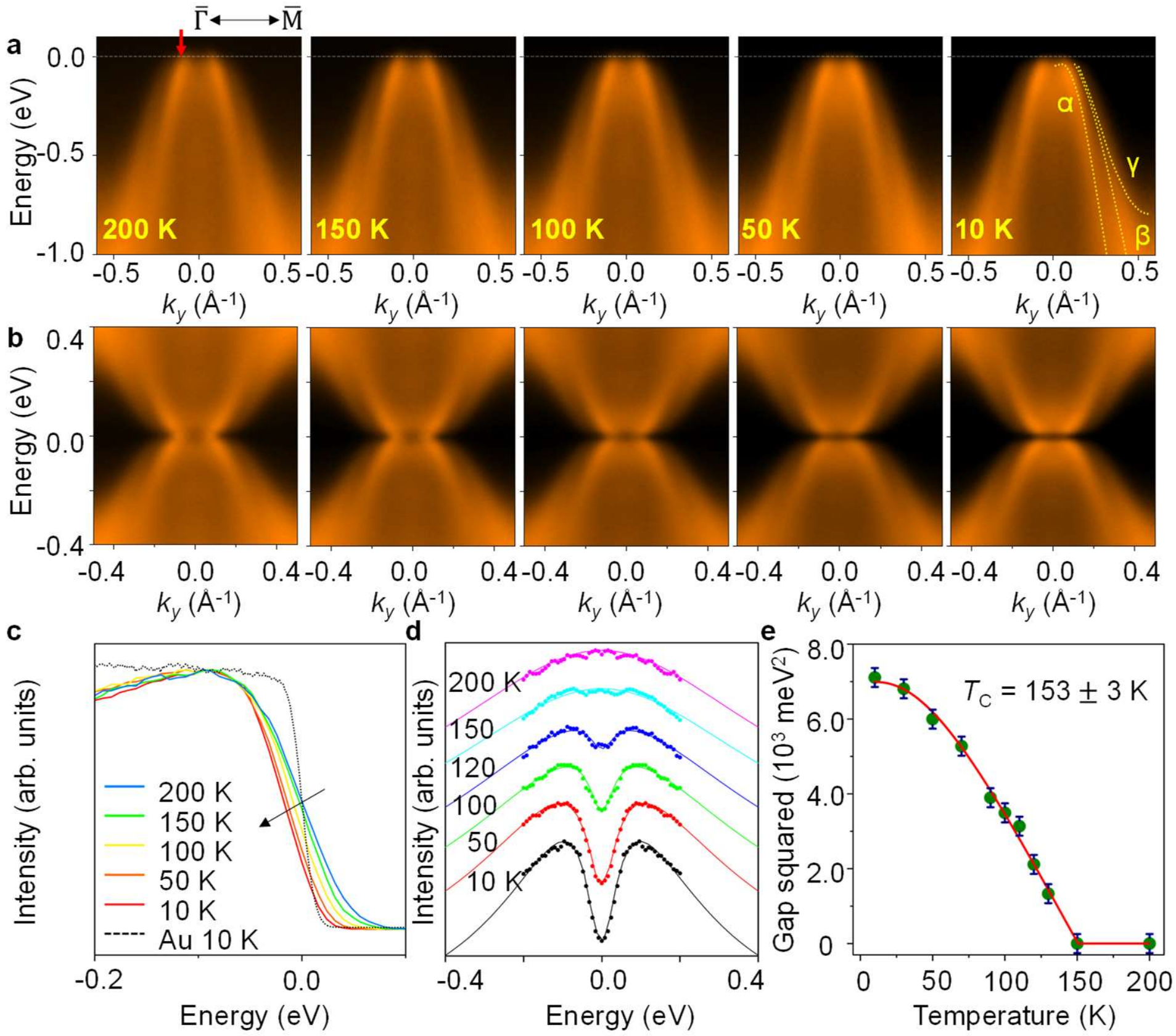

**Figure 3**. The tunneling spectral gap and temperature dependent LEED patterns for $Cu_xS$ films. (a) d*I*/d*V* spectra taken at the interior of the $Cu_xS$ islands at 3.5 K are superimposed on top of the ARPES spectra taken along the $\overline{\Gamma M}$ direction. The arrows indicate the valence and conduction band positions. Below the Fermi level, the valence band position matches well with the ARPES data. (b) Evolution of $Cu_xS$ (annealed at 330 ºC) diffraction spots with varying temperature. The electron incidence angle relative to surface normal is labeled. Two preferred orientations of $Cu_xS$ domains are revealed and no notable changes are observed across the transition.

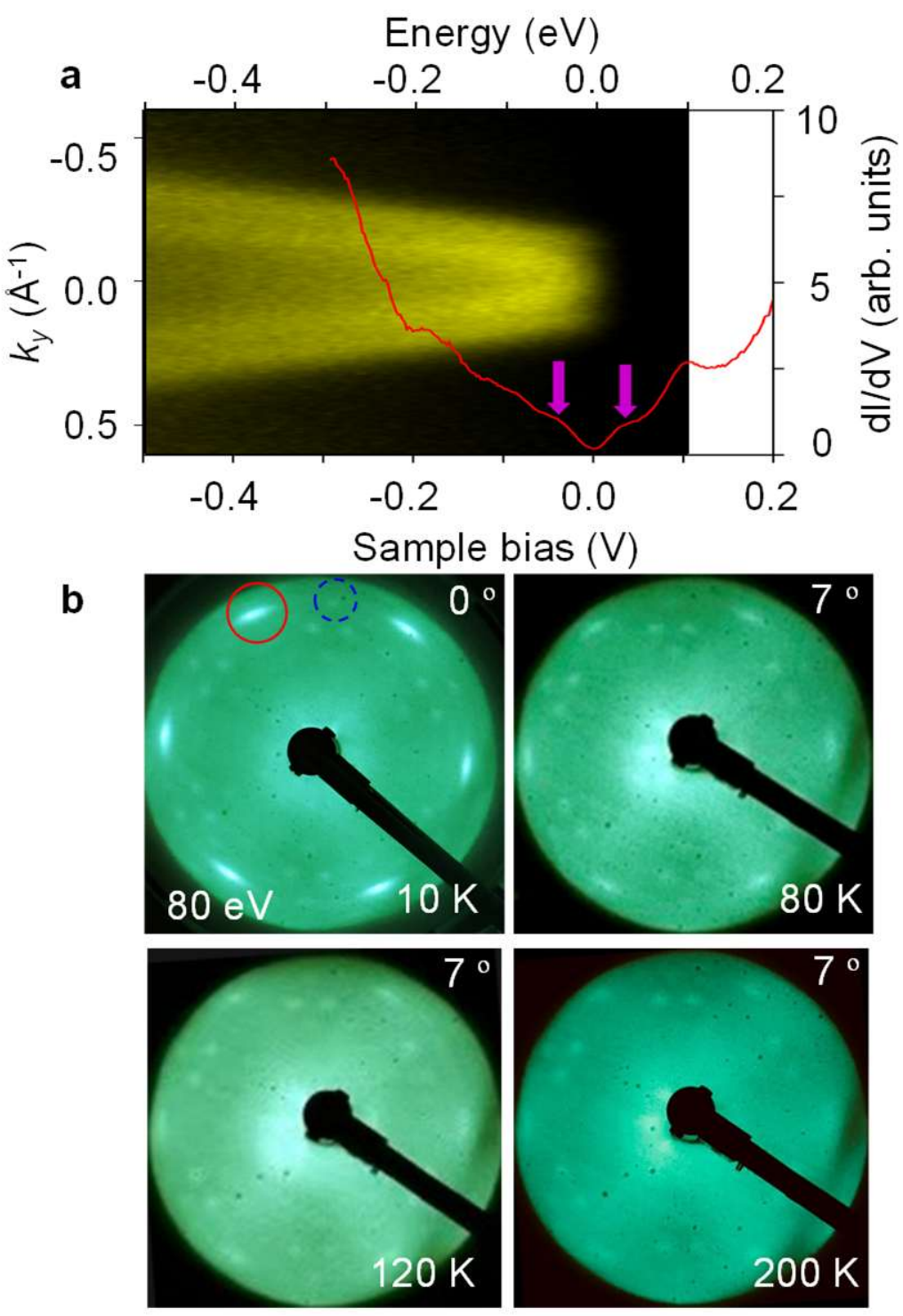

**Figure 4**. Phase diagram for $Cu_xS$. Transition temperature as a function of hole carrier density (*p*) around the Fermi level. Schematic diagrams of the electronic structure for different carrier densities are shown to demonstrate the relation of the band structure of $Cu_xS$ with the phase transition. Two primary hole-like bands were shown for clarity. The grey dashed line in each diagram denotes the Fermi level.

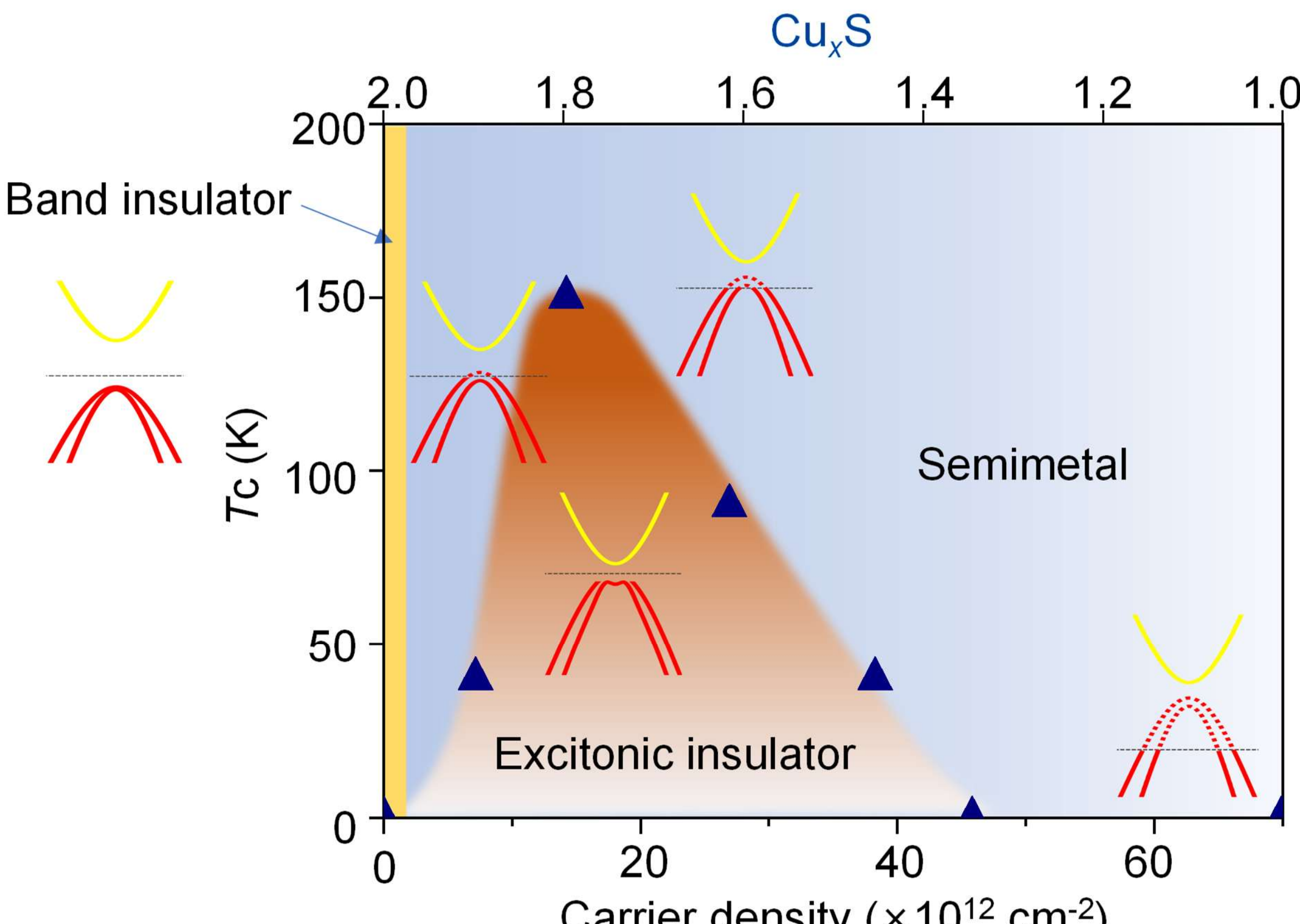